\documentclass[aps,twocolumn,superscriptaddress]{revtex4}
\input epsf

\begin{document}

\title{Fragile-to-strong transition and polyamorphism in the energy
landscape of liquid silica}

\author{Ivan Saika-Voivod}
\affiliation{Department of Applied Mathematics,
University of Western Ontario, London, Ontario N6A~5B7, Canada}

\author{Peter H. Poole}
\affiliation{Department of Applied Mathematics,
University of Western Ontario, London, Ontario N6A~5B7, Canada}

\author{Francesco Sciortino}
\affiliation{Dipartimento di Fisica and Istituto Nazionale per la
Fisica della Materia, Universita' di Roma La Sapienza, Piazzale Aldo
Moro~2, I-00185, Roma, Italy}

\date{\today}

\begin{abstract}
We study the static and dynamic properties of liquid silica over a
wide range of temperature $T$ and density $\rho$ using computer
simulations.  The results reveal a change in the potential energy
landscape as $T$ decreases that underlies a transition from a fragile
liquid at high $T$ to a strong liquid at low $T$.  We also show that a
specific heat anomaly is associated with this fragile-to-strong
transition, and suggest that this anomaly is related to the
polyamorphic behaviour of amorphous solid silica.
\end{abstract}

\maketitle

Liquid silica is the archetypal glass former, and compounds based on
silica are ubiquitous as natural and man-made amorphous materials.
Liquid silica is also the extreme case of a ``strong liquid,'' one
having a variation of viscosity $\eta$ with temperature $T$ that
closely follows the Arrhenius law $\log \eta\sim 1/T$ as the liquid is
cooled toward its glass transition temperature
$T_g$~\cite{A91,richet}.  In contrast, most liquids are to some degree
``fragile,'' exhibiting significantly faster-than-Arrhenius increases
of $\eta$ as $T \to T_g$.  Recent studies focusing on the properties
of the potential energy hypersurface (PES)---or ``energy
landscape''---of the liquid have demonstrated the controlling
influence of the PES on transport properties as $T\to
T_g$~\cite{SDS98,SKT99,BH99,S01,caax}.  However, these studies have
only addressed fragile liquids, and the origin of strong liquid
behaviour in terms of the energy landscape has not yet been resolved.
It is this question that we address here.

In a molecular dynamics computer simulation of an equilibrium liquid,
the diffusion coefficient $D$ is readily evaluated from the particle
trajectories. Like $\eta$, $D$ is a characteristic transport property
whose deviations from the Arrhenius law serve to classify a liquid as
strong or fragile.  The theory of Adam and Gibbs (AG)~\cite{AG65}
states that $D$ is related to the configurational entropy, $S_c$ via,
\begin{equation}
D=D_0 \exp(-A/TS_c),
\label{AG}
\end{equation}
where the parameters $D_0$ and $A$ are commonly assumed to be
$T$-independent.  $S_c$ quantifies the number of distinct
configurational states explored by the liquid in equilibrium.  These
states have been proposed to correspond to the ``basins'' of the PES
sampled by the liquid~\cite{sw,S95}.  A basin is the set of points in
phase space representing configurations having the same local minimum.
The local minimum configuration is termed an inherent structure (IS),
and is identified in simulation by a steepest descent minimization of
the potential energy.

Following the IS thermodynamic formalism of Stillinger and
Weber~\cite{sw}, the internal energy of the liquid can be expressed as
$E= e_{IS} + E_{harm} + E_{anh}$, where $e_{IS}$ is the average IS
energy and the last two terms are the average contributions to $E$ due
to thermal excitations about the IS.  $E_{harm}$ is the average
harmonic contribution determined from a quadratic approximation to $E$
around each IS minimum, while $E_{anh}$ is the remaining, necessarily
anharmonic contribution.  The harmonic and anharmonic potentials
characterize the shape of the basin.

If the shape of the basins does not depend on $e_{IS}$ (a condition
satisfied at constant density $\rho$ in the present study; see
Methods), then $S_c$ can be calculated along an isochore by
integrating the $T$ dependence of $e_{IS}$ at constant volume
$V$~\cite{SKT99}:
\begin{equation}
S_{c} = S_c^0 + \int_{T_0}^T \frac{1}{T'}
\left ( \frac{\partial{e_{IS}}}{\partial{T'}} \right )_V dT', 
\label{sconf}
\end{equation}
where $S_c^0$ is the value of $S_c$ at a reference $T=T_0$ (see
Methods).  Eq.~\ref{sconf} highlights that the $T$ dependence of $S_c$
arises solely from changes in $e_{IS}$~\cite{sw,SKT99}.  Evaluation of
$S_c$ when the basin shape depends on $e_{IS}$ is also
feasible~\cite{FD,starr,S01}.

For a strong liquid that satisfies Eq.~\ref{AG}, the $T$ dependence of
$S_c$ must approach a constant to recover Arrhenius behaviour.  From
Eq.~\ref{sconf}, if $S_c$ is constant, then so must be the variation
of $e_{IS}$ with $T$.  This behaviour would be qualitatively different
from that found in simulations of fragile liquids.  For example,
recent studies of a binary Lennard-Jones liquid have shown that
$e_{IS}\propto -1/T$, a dependence that results from a Gaussian
distribution of IS energies~\cite{BH99,S01}.  When the IS energy
distribution is Gaussian, fragility has been shown to depend on the
total number of IS states, the width of the Gaussian, and the
dependence of the basin shape on $e_{IS}$~\cite{S01,SPEEDY}.  However,
it is not known if the PES of a strong liquid is similarly
characterized by a Gaussian distribution of IS energies but with
parameters that differ from the fragile liquid case, or if the PES is
qualitatively different from that of a fragile liquid.

In addition, some analyses of experimental data for silica suggest
that the liquid may be fragile at very high $T$~\cite{HDR96,RHN98}.
Recent simulations~\cite{HK99} of the BKS~\cite{BKS90} model of silica
between $T=2750$ and $6000$~K have also shown that at the onset of
slow dynamics, as reflected for example by the presence of two-step
relaxation in structural autocorrelation functions, BKS silica is a
fragile liquid.  At about 3300~K BKS silica transforms to a strong
liquid and the $T$-dependence of all characteristic relaxation times
becomes Arrhenius~\cite{HK99}.  The relationship of the PES to such a
``fragile-to-strong'' transition is not yet known.

To address these questions, we conduct extensive equilibrium
simulations of BKS silica over a large range of $T$ and $\rho$ to
examine the fragile-to-strong transition, and to identify the energy
landscape behaviour that underlies it.  The results clarify our
understanding not only of the origins of silica's status as a strong
liquid, but also of the dynamical behaviour of a wider class of
liquids that are to some degree silica-like, most notably water and
silicon~\cite{ssx}, but also other systems such as
BeF$_2$~\cite{caabef2}.  Indeed, the concept of a
``fragile-to-strong'' transition was first proposed for the case of
deeply supercooled water~\cite{angellwaterII}.

\begin{figure}
\hbox to\hsize{\epsfxsize=1.0\hsize\hfil\epsfbox{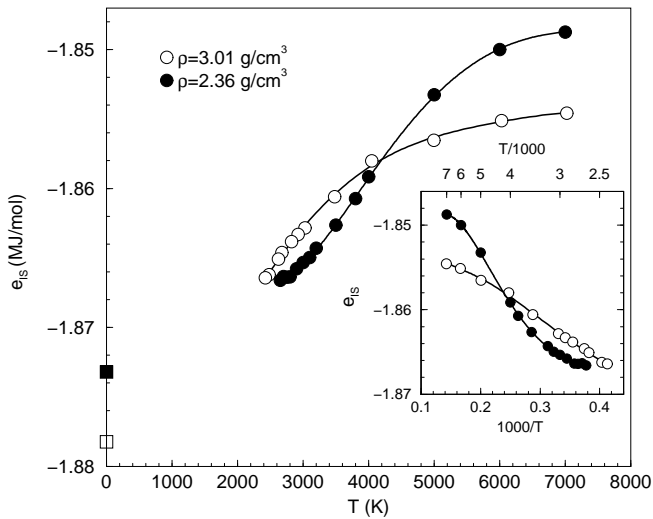}\hfil}
\caption{$e_{IS}$ as a function of $T$ along two isochores.  At $T=0$
we show the energy $E_0$ of the crystalline system at
$\rho=2.36$~g/cm$^3$ (filled square) and $\rho=3.01$~g/cm$^3$ (open
square).  $E_0$ is found by calculating the $V$ dependence of the
potential energy at $T=0$ of three crystal polymorphs of silica
(stishovite, coesite and quartz), and then using the common tangent
construction to determine the potential energy of the heterophase of
coexisting crystals that would be the ground state at the required
bulk value of $\rho$.  Inset: $e_{IS}$ for the same isochores as in
the main panel, plotted as a function of $1/T$.  Only the high density
data show a clear $1/T$ behaviour at low $T$.}
\label{eis}
\end{figure}

Fig.~\ref{eis} shows the $T$ dependence of $e_{IS}$ along two
isochores.  The shape of the higher $\rho$ isochore is similar to that
found for fragile liquids.  However, at the lower $\rho$---which is
close to the experimental $\rho$ at ambient pressure---$e_{IS}$
exhibits an inflection, passing from concave downward at high $T$ to
concave upward at low $T$.  Fig.~\ref{eis} also shows the potential
energy $E_0$ at $T=0$ of the corresponding equilibrium crystalline
system for the same $\rho$.  Since the energy of the lowest IS cannot
be less than $E_0$, $E_0$ sets a lower bound on $e_{IS}$.  The value
of $E_0$ relative to the measured $e_{IS}$ curves confirms that an
inflection occurs.  

The inset of Fig.~\ref{eis} shows that the relation $e_{IS}\propto
-1/T$, the hallmark of a Gaussian distribution of IS energies, is not
obeyed along our lower $\rho$ isochore.  The breakdown of this
relation does not arise from changes in the shapes of the basins
sampled at different $T$.  Indeed, the distribution of curvature of
the PES around the IS (the vibrational density of states) is found to
be independent of the basin depth for all the isochores we study.
Hence, within the harmonic approximation, all basins of the same
$\rho$ are characterized by the same vibrational entropy.

\begin{figure}
\hbox to\hsize{\epsfxsize=1.0\hsize\hfil\epsfbox{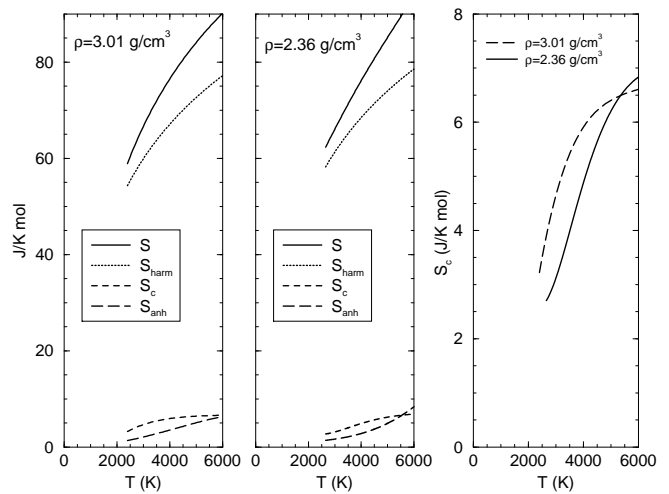}\hfil}
\caption{$S$ and its component contributions as a function of $T$ at
$\rho=3.01$~g/cm$^3$ (left panel) and $\rho=2.36$~g/cm$^3$ (centre
panel). $S_c$ as a function of $T$ along two isochores (right panel).}
\label{sc}
\end{figure}

\begin{figure}
\hbox to\hsize{\epsfxsize=1.0\hsize\hfil\epsfbox{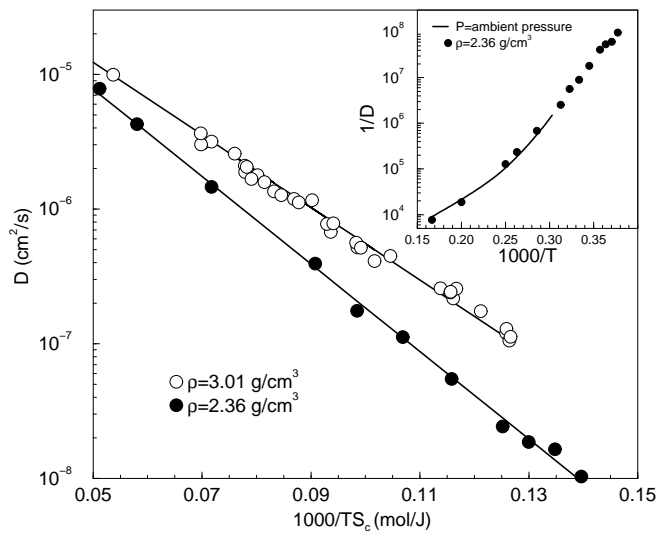}\hfil}
\caption{$\log D$ versus $1/TS_c$ along two isochores.  Inset:
Arrhenius plot of $1/D$ along the ambient pressure isobar, found by
interpolating our isochoric data; and along the $\rho=2.36$~g/cm$^3$
isochore.  Note the non-Arrhenius to Arrhenius crossover observed both
at constant $\rho$ and at constant $P$.  All $D$ values reported here
are for Si atoms.}
\label{AGfig}
\end{figure}

Like $e_{IS}$, $S_c$ (Eq.~\ref{sconf}) also exhibits an inflection
along our lower $\rho$ isochore (Fig.~\ref{sc}).  In the same range of
$T$, we find (Fig.~\ref{AGfig}) that along each isochore $D$ satisfies
the AG relation within numerical error.  That is, despite the change
in the nature of the $T$ dependence of $S_c$, the transport properties
of the liquid adjust (from non-Arrhenius to Arrhenius) so as to
maintain the AG relation, a finding that demonstrates of the
robustness of Eq.~\ref{AG}~\cite{robinAG4HS,SSNSS00}.  The low $\rho$
data thus reveal the signature in the energy landscape of a
fragile-to-strong transition.  The rapid variation of landscape
properties at high $T$ corresponds to a fragile regime.  As $T$
decreases, the inflections of $e_{IS}$ and $S_c$ mark the onset of a
regime in which the rate of change of these quantities decreases,
consistent with the system's approach to the strong liquid limit.

We note that for the present model, $S_{c}$ is much smaller than the
vibrational entropy ($S_{harm}+S_{anh}$) and of the same order as
$S_{anh}$.  Moreover, we find that for BKS silica the crystalline
vibrational entropy differs significantly from the liquid vibrational
entropy, as also found by Richet~\cite{richet} in his extensive
analysis of silicate melts. This confirms that for silica, the
identification of the excess entropy (liquid entropy minus crystal
entropy) with $S_c$ may not be adequate~\cite{Johari,goldstein}.  This
highlights the value of finding $S_c$ via the ``all-liquid
route''~\cite{coluzzi} implemented here.

\begin{figure}
\hbox to\hsize{\epsfxsize=1.0\hsize\hfil\epsfbox{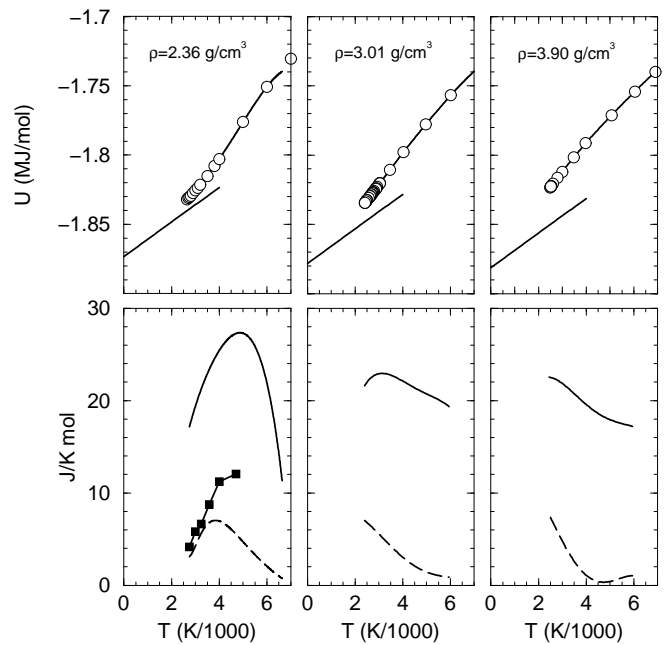}\hfil}
\caption{Upper panels: Isochores of the liquid potential energy $U$.
Lower panels: $C_V-(3/2)R$ (solid lines) and $C_V^{IS}$ (dashed lines)
as a function of $T$.  In the upper panels are shown lines of slope
$\frac{3}{2} R$ whose values at $T=0$ are the potential energies of
the corresponding crystalline systems, calculated as described in
Fig.~\ref{eis}; these lines are lower bounds for $U$ of the liquid.
In the left-most lower panel are shown (filled squares) estimates of
the configurational part of $C_V$ recently evaluated by Scheidler,
et~al.~\cite{kobcv}, calculated from the time dependence of the
temperature fluctuations.}
\label{E}
\end{figure}

The influence of the energy landscape is sufficiently prominent to
appear in the total thermodynamic properties~\cite{sriISKT}.  The
isochoric specific heat $C_V$ can be written as
$C_V=C_V^{IS}+3R+C_V^{anh}$, where each term is the derivative with
respect to $T$ of the corresponding contribution to $E$, and $R$ is
the gas constant. The inflection in the $T$ dependence of $e_{IS}$
corresponds to a maximum in $C_V^{IS}$ (Fig.~\ref{E}) that is the
origin of a $C_V$ anomaly, in the form of a peak, in the interval of
$T$ corresponding to the fragile-to-strong transition.  An analogous
$C_V$ anomaly has recently been predicted to occur in the silica-like
liquid BeF$_2$~\cite{caabef2}, and in theoretical models designed to
reproduce a fragile-to-strong transition~\cite{jagla}.  High $T$
experiments that test for this anomaly, though challenging, can thus
directly seek the thermodynamic signature of the fragile-to-strong
transition in these systems.

The peak we find in $C_V$ occurs at $T$ near the temperature of
maximum density of BKS silica, and in the vicinity of a maximum of the
isothermal compressibility $K_T$ predicted for this
model~\cite{SSP00}. Thermodynamic anomalies, such as peaks in $C_V$
and $K_T$, have been associated with the physics of liquid-liquid
transitions and with polyamorphism in glasses, i.e. the abrupt
pressure-induced transition of a low-density glass to a high-density
glass~\cite{sci}.  Polyamorphism is observed experimentally during
compression of silica glass, and the $C_V$ maximum found here may be
the thermal anomaly corresponding to polyamorphism's mechanical
anomaly.  Along different isochores we find that the $T$ at which the
$C_V$ maximum occurs decreases with increasing $\rho$, as would be
expected for an anomaly related to polyamorphism in silica.

These observations together suggest that the present fragile-to-strong
transition is the dynamical transition corresponding to the
thermodynamic crossover in the liquid that presages
polyamorphism~\cite{jagla}.  More generally, our results provide a
basis for considering all strong liquids as candidates for
polyamorphism: a strong liquid arises via a fragile-to-strong
transition, associated with which may be thermodynamic anomalies that
are the liquid-state precursors to polyamorphism.

\medskip
\noindent
{\bf Appendix: Methods}

\medskip
\noindent
{\it Computer simulations:} Results are based on molecular dynamics
simulations of BKS silica.  All data are obtained from systems of
$N=1332$ atoms (888 O and 444 Si atoms), except for the
$\rho=2.36$~g/cm$^3$ isochore, where we employ 999 atoms to facilitate
equilibration at low $T$. At $\rho=2.36$~g/cm$^3$ 8 independent runs
for each state point are performed.  All data reported here are for
liquid states in equilibrium.  Equilibration is confirmed by ensuring
that runs are longer than the slowest relaxation time in the system as
evaluated from the collective (coherent) density-density correlation
function.  The lowest $T$ runs exceed $80$~ns.  Simulations are
carried out in the constant-$(N,V,E)$ ensemble, and long-range
electrostatic interactions are accounted for by Ewald summation.  We
evaluate $e_{IS}$ by conducting conjugate gradient minimizations of up
to 1000 equilibrium liquid configurations and averaging the results.

\medskip
\noindent
{\it Evaluation of the total entropy:} For a given $\rho$, $S$ at
$T=T_0=4000$~K is calculated using thermodynamic integration: We first
find the free energy difference between an ideal gas and a binary
mixture Lennard-Jones (LJ) system at the chosen $\rho$ by integrating
the excess pressure along an isotherm from the high $V$ limit where
the system behaves as an ideal gas.  We then carry out a set of
simulations at constant $V$ and $T$ that continuously convert the LJ
system to BKS silica, by employing a hybrid potential $\phi=\lambda
\phi_{BKS} + (1-\lambda)\phi_{LJ}$~\cite{mezei}.  An appropriate
thermodynamic integration from $\lambda=0$ to $1$ yields the free
energy of BKS silica, from which $S$ at $T_0$ and the chosen $\rho$ is
calculated.  $S$ at other $T$ is found by further thermodynamic
integration at constant $V$.

\medskip
\noindent
{\it Evaluation of the vibrational entropy:} We evaluate $S_{harm}$
from the spectrum of eigenfrequencies $\nu$ (i.e. the vibrational
density of states) calculated from the IS's at each $T$, using
$S_{harm}=k\sum_{i=1}^{3N}[1-\log(h \nu_i/kT)]$, where $k$ and $h$ are
Boltzmann's and Planck's constants, respectively.  To obtain $S_{anh}$
we use $E_{anh}$.  We evaluate $E_{anh}=E-E_{harm}-e_{IS}$ and then
fit $E_{anh}$ with a polynomial constrained to be zero and have zero
slope at $T=0$.  When the shape of the basins does not depend on
$e_{IS}$, $S_{anh}$ may be calculated by thermodynamic integration
using the fitted form of $E_{anh}$ from $T=0$ to the desired $T$.  In
terms of the above quantities, the integration constant in
Eq.~\ref{sconf} is thus $S_c^0=S(T_0) - S_{anh}(T_0) - S_{harm}(T_0)$.

\medskip
\noindent{\it Isochoric invariance of basin shape:} Our assumption
that the shape of the basins is independent of $e_{IS}$ along an
isochore is based on two observations: (i) We find that the
vibrational density of states (the $\nu$ spectrum) is independent of
$e_{IS}$ along an isochore.  (ii) The anharmonic energy of IS
configurations heated from $T=0$ to a chosen $T$ follows the
$E_{anh}=E-E_{harm}-e_{IS}$ curve calculated from equilibrium
simulations.  This is only possible if the anharmonic character of the
basins is also independent of $e_{IS}$.

\begin{acknowledgments}
We thank C.A.~Angell, W.~Kob, S.~Sastry and R.~Speedy for useful
discussions.  ISV and PHP acknowledge NSERC (Canada) for financial
support, and the Compaq-Western Centre for Computational Research for
computing resources. FS acknowledges support from the INFM
``Iniziativa Calcolo Parallelo,'' PRE-HOP and from MURST PRIN 2000.
\end{acknowledgments}

\end{document}